\documentclass[fleqn,usenatbib]{mnras}
\usepackage[utf8]{inputenc}
\usepackage{csquotes}
\usepackage{bm}
\usepackage{amsmath}
\usepackage{mathtools} 
\usepackage{amssymb}
\usepackage{graphicx}
\usepackage[usenames,dvipsnames]{color}
\usepackage{hyperref}
\usepackage{widetext}
\hypersetup{
    colorlinks=true,
    linkcolor=blue,
    filecolor=magenta,      
    urlcolor=cyan,
    pdftitle={Overleaf Example},
    pdfpagemode=FullScreen,
    }
\let\oldhat\hat
\renewcommand{\hat}[1]{\oldhat{\mathbf{#1}}}

\usepackage{bm}

\title[]{Constraining cosmological vorticity modes with CMB secondary anisotropies}

\author[W.~R.~Coulton et al.]{William~R.~Coulton,$^1$ 
Kazuyuki~Akitsu,$^2$ Masahiro~Takada$^3$\\
$^1$Center for Computational Astrophysics, Flatiron Institute, 162 5th Avenue, New York, NY 10010, USA\\
$^2$School of Natural Sciences, Institute for Advanced Study, 1 Einstein Drive, Princeton, NJ 08540, USA\\
$^3$Kavli Institute for the Physics and Mathematics of the Universe (WPI), UTIAS, \\
The University of Tokyo,  5-1-5 Kashiwanoha, Kashiwa, Chiba 277-8583, Japan
}

\date{\today}

\begin{document}

\maketitle

\begin{abstract}
Observational searches for large-scale vorticity modes in the late time Universe are underexplored. Within the standard $\Lambda$CDM model, this is well motivated given the observed properties of the cosmic microwave background (CMB). However, this means that searches for cosmic vorticity modes 
can serve as a powerful consistency test of our cosmological model.
We show that through combining CMB measurements of the kinetic Sunyaev-Zel'dovich and the moving lens effects with galaxy survey data we can constrain vorticity fields independently from the large scale cosmic velocity field. 
This approach can provide stringent constraints on the largest scale modes and can be achieved by a simple change in the standard estimators.  Alternatively if one assumes there are no cosmic vorticity modes, this estimator can be used to test for systematic biases in existing analyses of kinetic Sunyaev-Zel'dovich effect in a manner analogous to curl-lensing.
\end{abstract}

\section{Introduction}
\label{sec:introduction}

Consistent measurements of the properties of the Universe across 13 billion years has established the  $\Lambda$CDM cosmological model \citep{Planck_2018_cosmoParams,Abbott_2022,dAmico_2021,2022PhRvD.105h3517K,Philcox_2022,Brout_2022}.  However, we still lack insight into the two of the key components, dark matter and dark energy, and there are now hints of potential cracks in this model -- in the form of tensions between probes of different eras 
\citep[see][for an overview of the `Hubble' and `$\sigma_8$' tensions]{Abdalla_2022,Schoneberg_2022}.
Testing the $\Lambda$CDM model through ever more powerful and varied tests is thus essential if we are to advance our understanding on these fundamental issues. 

Perturbations to the homogeneous, isotropic background Friedmann–Lemaître–Robertson–Walker (FLRW) Universe can be decomposed into three types of perturbations: scalars, vectors and tensors. At the linear level, these three  are decoupled and do not mix \citep[see e.g.][]{2020moco.book.....D,Weinberg_2008,Mukhanov_2005}. Through measurements of the cosmic microwave background (CMB), we have discovered that the early Universe is dominated by small scalar perturbations. As linear vector and tensor decay within the standard $\Lambda$CDM model, these high redshift measurements place tight bounds on the expected level of vector and tensor modes \citep{Bennett_2003,Ichiki_2012,Saga_2014}. Whilst the evolution of the small, scalar perturbations is non-linear, these processes do not generate large levels of large-scale vector and tensor perturbations \citep{2007PhRvD..76h4019B,Lu_2009,Pueblas_2009,Hahn_2015,Jelic-Cizmek_2018}. Thus an important prediction of our current model is the absence of late-time, large-scale, cosmic vorticity, velocity fields that cannot be expressed as the gradient of a scalar and are a type of vector perturbation. To date, there have been only a few analyses of vorticity/vector modes in the late time Universe - \citet{2021NatAs...5..283M} search for vorticity modes in galaxy spins, whilst \citet{Namikawa_2013} use the CMB lensing curl mode. 

Recently numerous authors have shown how measurements of the kinetic Sunyaev-Zel'dovich effect (kSZ) or moving lensing (ML) effect, two CMB secondary anisotropies imprinted on the CMB as photons propagate to the observer from the surface of last scattering, can be used to measure the large scale velocity field \citep{Li_2014,Alonso_2016,Terrana_2017,smith2018ksz,Yasini_2019,Hotinli_2019,McCarthy_2020}. This achieved via a technique called ``kSZ/ML tomography" where high resolution, low noise measurements of the CMB are combined with measurements of the large scale structure of the Universe, such as galaxy surveys. This technique is a highly promising avenue for studying large scale velocity fields with upcoming surveys predicted to make ${\cal O}(100)$s signal-to-noise ratio (SNR) measurements of velocity fields on $k\lesssim 10^{-3}~h$Mpc$^{-1}$ \citep{smith2018ksz,Munchmeyer_2018}.

In this work we explore whether kSZ and ML measurements can be used to study late-time vorticity modes. Such measurements would provide strong tests of the standard model expectation that these modes are small, constrain novel cosmological models that can produce large, late-time vorticity modes or, if one assumes there are no vorticity modes, act as a systematics-null test for kSZ measurements. This work complements recent work by \citet{Bonvin_2018}, which explores how redshift space distortions can be used to constrain vorticity modes, and we find that our approach offers a significantly more powerful means of constraining the largest scales.

Throughout this paper, we work in the `snapshot' geometry described in \citet{smith2018ksz} where we compute our observables by treating the universe as a periodic box at a distance $\chi_*$ from the observer. This setup allows the intuition to be clearly developed and the results can be extend to include light-cone effects and a treatment of the curved sky in an analogous way to \citet{Terrana_2017}. In Appendix \ref{app:lightConeGeom} we outline how this idea can be framed in the framework of \citet{Terrana_2017}.

This paper is strucutred as follows: in Section \ref{sec:vorticityModes} we review vorticity modes. In Sections \ref{sec:kSZeffect} and \ref{sec:MLeffect}
we show how vorticity modes source kSZ and ML anisotropies. In Section \ref{sec:vel_recon}, we describe how to use these effects to reconstruct the vorticity perturbations and present a forecast of this approach in Section \ref{sec:constraints}. Our conclusions and outlook are presented in Section \ref{sec:conclusions}.
\section{Vorticity modes}\label{sec:vorticityModes}
An arbitrary vector can be decomposed with Helmholtz decomposition into a part that is curl-free, $u$, and a part that is divergence-free, $\omega$. Thus, we can decompose the Fourier space velocity perturbations into divergence, $u(\mathbf{k})$, and curl (vorticity), $\omega_i(\mathbf{k})$ contributions as  
\begin{align}
\mathbf{v}_i = \frac{i k_i}{k} u(\mathbf{k}) + i\omega_i(\mathbf{k}),
\end{align}
where the curl component satisfies $\omega_i k_i = 0$. To proceed it is useful to represent the divergence-free mode with the polarization vectors
\begin{align}
\omega_i(\mathbf{k}) = \sum_{s}\epsilon^{s}_i(\hat{\mathbf{k}}) \omega^{(s)}(\mathbf{k}),
\end{align}
where $\epsilon^s_i$ are the polarization basis vector  and $s$ denotes the two polarization states $(+1,-1)$. It is convenient to work in a rotated basis defined as 
\begin{align}
\omega^{\text{+}} = \frac{1}{\sqrt{2}}\left[\omega^{(+1)}+\omega^{(-1)}\right]\, \text{ and } \, 
\omega^{\times} = \frac{1}{\sqrt{2}}\left[\omega^{(+1)}-\omega^{(-1)}\right].
\end{align}
In linear cosmological perturbation theory curl modes are not sourced by density (scalar) modes and so are expected to vanish. This means that these curl modes are a powerful probe of new physics. As an example consider cosmological vector perturbations. In the synchronous gauge, the perturbed FLRW metric can be written as
\begin{align}
&\mathrm{d}s^2=a(\tau)^2 \times \nonumber \\ &\left(-\mathrm{d}\tau^2+\left[(1-2\phi)\delta_{ij}+2(D_{,ij}+\Omega_{(i,j)} +h^{TT}_{ij} ) \right]\mathrm{d}x^i\mathrm{d}x^j \right),
\end{align}
where $\phi(\mathbf{x},t)$ and $D(\mathbf{x},t)$ are scalar perturbations, $\Omega_i(\mathbf{x},t)$ is the transverse vector perturbation and $h_{ij}^{TT}(\mathbf{x},t)$ is the transverse, traceless tensor perturbation.  The vector modes source vorticity perturbations as
\begin{align}
\omega_{i}(\mathbf{x},\tau) = -\frac{\nabla^2 \Omega'_i}{16\pi Ga^2(\epsilon_0+p_0)},
\end{align}
where $a$ is the scale factor, $G$ is the Gravitational constant, $'$ denotes the conformal time derivative, and $\epsilon_0$ and $p_0$ are the energy density and pressure. 
Note that, in matter domination 
 and without a source, the vector perturbations, $\Omega_i$, decay as $a^{-\frac{5}{2}}$ and so their induced vorticity decays as, $a^{-1}$.

To have observable vector modes they thus need to be sourced. In general vector modes can be sourced whenever there is non-vanishing anisotropic stress such as neutrino velocity isocurvature modes \citep{Lewis_2004a} or primordial magnetic field \citep{Lewis_2004b}. More exotic sources include modifications to general relativity \citep{Battye_2017}, topological defects \citep{Seljak_1997,Durrer_2002,Daverio_2016}, a global rotation \citep{Carneiro_2000} and some dark energy models \citep{Battye_2006,Battye_2007,Battye_2009}. 

\section{Kinetic Sunyaev-Zel'dovich effect}\label{sec:kSZeffect}
The kinetic Sunyaev-Zel'dovich (kSZ) effect occurs when CMB photons are Thomson scattering off electrons that are moving with  
bulk motions, such as in galaxy groups and clusters \citep{Sunyaev_1980}. This scattering generates anisotropies in the CMB, without altering the CMB blackbody spectrum, that depend on the electron's properties via
\begin{equation}
    \frac{\Delta T^{\rm kSZ}(\hat{n})}{T_{\rm CMB}} = - \int{\mathrm{d}\chi a(\chi) \sigma_T n_e(\chi) \mathbf{v} (\chi \hat{n}) \cdot \hat{n}  e^{-\tau(\chi)}},
    \label{eq:kSZ} 
\end{equation}
where $\bf{v} $ is the electron velocity, $n_e$ is the electron density, $\chi$ is the comoving distance, $\sigma_T$ is the Thomson scattering cross-section and $\tau$ is the optical depth. In our simplified geometry the line-of-sight integral integrates across our box and the observed angle $\mathbf{n}$ describes points of the surface of the box at location $(x,y) = \chi \mathbf{n} $.

In Fourier space we can see that the kSZ anisotropies are given by
\begin{align}\label{eq:kSZ_allSources}
 \tilde{T}^{\rm kSZ}(\ell) =& i\frac{a \sigma_T \bar{n}_e e^{-\tau}}{\chi_*^2}\int \frac{\mathrm{d}k_1^3}{(2\pi)^3}\frac{\mathrm{d}k_2^3}{(2\pi)^3}(2\pi)^3
 \delta_D^{(3)}\!\!\left(\frac{\boldsymbol{\ell}}{\chi_*}-\mathbf{k}_1-\mathbf{k}_2\right) \nonumber \\ &
\times \delta_e(\mathbf{k}_1)\left [\cos\theta_{k_2} u(\mathbf{k}_2)-\sin\theta_{k_2}\omega^+(\mathbf{k}_2) \right],
\end{align}
where $\delta^{(3)}_D({\bf x})$ is the 3D Dirac delta function and $\delta_e(\mathbf{k})$ is the Fourier transform of the electron density fluctuation field. 
The scalar and vector sources contribute via different angular dependencies, implying that these effects can be separated. Further note that the kSZ effect is only sensitive to the $\omega^+$ modes.

\section{The moving lens effect}\label{sec:MLeffect}
The moving lens effect is a second-order effect that generates anisotropies in the CMB. This effect can be understood from two equivalent perspectives either as part of the Rees-Scamia, or non-linear integrated Sachs-Wolfe effect, or as lensing by a moving cluster \citep{1968Natur.217..511R,Birkinshaw_1983,Gurvits_1986,Lewis_2006,Yasini_2019,Hotinli_2019}. The first perspective describes how the energy of photons is altered as they enter the potential of clusters moving perpendicular to the line of sight: if they pass in-front of the cluster they will be blueshifted as they fall into clusters however as they leave they are redshifted by a larger amount as the cluster has moved closer to the line of sight and deepened the potential. Note that there is no additional blue-shifting from the change in the potential as the cluster's velocity is perpendicular to the photon trajectory. The opposite effect occurs for photons passing behind the cluster. The second perspective can be understood by considering photon trajectories as viewed in the cluster rest frame. In this frame a dynamical dipole is seen and the dipolar distorted photons are deflected by the cluster's mass. Boosting back to the CMB rest frame removes the initial, unlensed dipole leaving a residual signal.

The size of the induced anisotropies is given by
\begin{align}
\frac{\Delta T^{\rm ML}(\mathbf{n})}{T_{\mathrm{CMB}} } = -2 \int \mathrm{d}\chi \mathbf{v}_{\perp} \cdot \nabla_{\perp}\Phi,
\end{align}
where $\Phi$ is the gravitational potential.  Inserting our expressions for the velocities into this, we find that the Fourier space ML anisotropies are given by
\begin{align}
&\tilde{T}^{\rm ML}(\boldsymbol{\ell})= \frac{2}{\chi_*^2} \int \frac{\mathrm{d}^3k}{(2\pi)^3}\Phi\!\left(\frac{\boldsymbol{\ell}}{\chi_*}-\mathbf{k}\right)\times \nonumber \\ & \Bigg[ u(\mathbf{k})\left( \frac{{\ell}}{\chi_*}\sin\theta_k\cos(\phi_k-\phi_\ell)-k \sin^2\theta_k\right) 
 \nonumber \\ &
+i\sin(\phi_k-\phi_\ell)
\omega^{\times}
(\mathbf{k}) \nonumber \\ &+ \omega^+(\mathbf{k})\left( \frac{{\ell}}{\chi_*} \cos\theta_k\cos(\phi_k-\phi_\ell)-k \sin\theta_k\cos\theta_k \right) \Bigg].
\end{align} 
There are several interesting features here: first the standard scalar velocity modes have a different angular dependence than the kSZ effect (this is expected as it probes transverse modes), which is essential for breaking the degeneracies with the vector kSZ modes. Second the vorticity modes likewise have a different angular dependence, and so in principle could be separated out from the other ML contributions. Finally the ML effect is also sensitive to the cross component ($\omega^\times$)
of the vector modes. However, this term is geometrically suppressed and as it is thus likely unmeasurable we do not consider it further here.

As was shown in \citet{Hotinli_2019} the moving-lens anisotropies are generally smaller than the kSZ anisotropies -- their contirubtion to the power spectrum at $\ell\sim 3000$ is an order of magnitude lower.

\section{Velocity Reconstruction}\label{sec:vel_recon}
Recent work 
\citep{Hand_2012,Terrana_2017,smith2018ksz,Cayuso_2021} has shown that by combining kSZ or ML anisotropies with galaxy position measurements large scale  scalar velocity modes, $ u(\mathbf{k} )$, can be reconstructed. The reconstruction uses a quadratic estimator
\begin{align} \label{eq:estimatorScalars}
\widehat{u}^{\rm X}(\mathbf{K}) = \int \frac{\mathrm{d}^3k\mathrm{d}^2\ell}{(2\pi)^2} W(\mathbf{k},\mathbf{l}) \tilde{T}^{\rm X}(\boldsymbol{\ell})\delta_g(\mathbf{k})
\delta^{(3)}_D\!\!\left(\mathbf{K}-\mathbf{k}-\frac{\boldsymbol{\ell}}{\chi_*} \right),
\end{align}
where $\delta_g(\mathbf{k})$ is the 3D Fourier transform of the galaxy density field, $\tilde{T}^{\rm X}(\boldsymbol{\ell})$ is the 2D Fourier transform of the kSZ or ML anisotropies and  $W(\mathbf{k},\mathbf{l})$ is a set of weights. This method is akin to CMB lensing where the lensing potential is reconstructed from measurements of the primary CMB anisotropies. Usually the weights in this estimator are chosen to obtain a minimum variance, unbiased estimator of the scalar velocity field i.e. $\langle \widehat{u}(\mathbf{k}) {u}^*(\mathbf{k'}) \rangle = (2\pi)^3 \delta^{(3)}(\mathbf{k}-\mathbf{k'}) P_{u}(k)$, where $P_u(k)$ is the velocity power spectrum. Given these requirements, the general case of reconstructing field $X$, the weights are \citep{darwish_2020}
\begin{align} \label{eq:wieghts}
 W(\mathbf{k},\mathbf{l}) & = \frac{{B^{X \delta_g T }}^*\!\!(-\mathbf{k}-\boldsymbol{\ell}/\chi_*,\mathbf{k},\boldsymbol{\ell}/\chi_*)}{P_{gg}^{\rm tot}(k) C_\ell^{\rm tot} P_{XX}(-\mathbf{k}-\boldsymbol{\ell}/\chi_*)} \times \nonumber \\ 
 &\hspace{-3em} \left[ \int \frac{\mathrm{d}^2\ell}{(2\pi)^2} \frac{{B^{X \delta_g T }}^*\!\!(-\mathbf{k}-\boldsymbol{\ell}/\chi_*,\mathbf{k},\boldsymbol{\ell}) B^{X \delta_g T }\!\!(-\mathbf{k}-\boldsymbol{\ell}/\chi_*,\mathbf{k},\boldsymbol{\ell}) }{P_{gg}^{\rm tot}(k) C_\ell^{\rm tot} P^2_{XX}\!\left(-\mathbf{k}-\boldsymbol{\ell}/\chi_*\right)} \right]^{-1 },
\end{align}
where ${B^{X \delta_g T }}^*\!(-\mathbf{k}-\boldsymbol{\ell}{/\chi_*},\mathbf{k},\boldsymbol{\ell}{/\chi_*})$ is the bispectrum between the true field, $X$, and the two fields in the quadratic estimator ($\delta_g$ and $T$), and $P_{gg}^{\rm tot}(k)$, $P_{XX}(k)$  and $C_\ell^{\rm tot}$ are the power spectra where the superscript ``tot" denotes that the power spectra include all contributions: signal, instrument noise, foregrounds and shot-noise -- as appropriate. For reconstructing scalar modes from the kSZ, this is $\langle u \delta_g T \rangle$ and is \citep{smith2018ksz}
\begin{align} \label{eq:bispecScalarkSZ}
B&^{u \delta_g T{\rm -kSZ}}\left(\mathbf{K},\mathbf{k},\frac{\boldsymbol{\ell}}{\chi_*}\right) = \nonumber \\ & \frac{a \sigma_T \bar{n}_e e^{-\tau}}{\chi_*^2}  i\cos\theta_K 
\left[{P_{uu}(K)}P_{ge}(k)-\frac{K}{k}{P_{ue}(K)}P_{ug}(k)\right].
\end{align}   
where $\bar{n}_e$ is the mean electron density at the scattering redshift, $P_{ge}$, $P_{ug}$ and $P_{ue}$ are the galaxy-electron, velocity-galaxy and velocity-electron power spectra. Similarly for scalar sources with the ML effect we have
\begin{align} \label{eq:bispecScalarML}
    B&^{u \delta_g T{\rm -ML}}\left(\mathbf{K},\mathbf{k},\frac{\boldsymbol{\ell}}{\chi_*}\right) = \frac{2}{\chi_*^2}
    \left[\frac{P_{uu}(K)}{K}P_{g\Phi}(k)+P_{u\Phi}(K)\frac{P_{gu}(k)}{k} \right] \nonumber \\& 
    \times \left[K\frac{{\ell}}{\chi_*}\sin\theta_K\cos(\phi_K-\phi_\ell)-K^2\sin^2\theta_K \right]. 
\end{align}
It is trivial to form a combined estimator by using the combined bispectra, i.e.
$ B^{u \delta_g T-{\rm Comb}} = B^{u \delta_g T-{\rm kSZ}}+B^{u \delta_g T-{\rm ML}} $.

Extending these estimators to reconstruct the vorticity modes is also straightforward. A quadratic estimator, $\hat{\omega}^+$, can be formed in an identical manner to Eq.~\ref{eq:estimatorScalars} with the vorticity bispectra used instead. For a reconstruction from kSZ anisotropies, this is
\begin{align} \label{eq:bispecVectorkSZ}
B&^{\omega^+ \delta_g T{\rm -kSZ}}\left(\mathbf{K},\mathbf{k},\frac{\boldsymbol{\ell}}{\chi_*}\right) =-\frac{a \sigma_T \bar{n}_e e^{-\tau}}{\chi_*^2}  i\sin\theta_K P_{eg}(k)P_{\omega\omega}(K) ,
\end{align}   
where $P_{\omega\omega}$ is the vorticity power spectrum and for reconstruction from ML anisotropies this is
\begin{align} \label{eq:bispecVectorkSZ}
B&^{\omega^+ \delta_g T{\rm -ML}}\left(\mathbf{K},\mathbf{k},\frac{\boldsymbol{\ell}}{\chi_*}\right) = P_{\Phi g}(k)P_{\omega\omega}(K) \nonumber \\ & 
\times \left[\frac{{\ell}}{\chi_*}\cos\theta_K \cos(\phi_K-\phi_\ell)-K\sin\theta_K\cos\theta_K \right].
\end{align}  
As for the scalar case a combined estimator can be formed by using the sum of the bispectra.

However, these vorticity estimators are not orthogonal to the scalar velocity estimators. This is problematic as measurements of the vorticity modes will likely be biased by the scalar modes, which are expected to be much larger than the vorticity terms. To avoid these biases we can construct vorticity estimators which are orthogonal, i.e. have zero response, to the scalar modes. 

The quadratic estimator that is unbiased, minimum variance and orthogonal to the scalar modes is given by
\begin{align}\label{eq:omega_estimator}
& \hat{\omega}^+(\mathbf{K}) = \int\frac{\mathrm{d}^2\ell}{(2\pi)^2} T(\boldsymbol{\ell})\delta_g\left(\mathbf{K}-\frac{\boldsymbol{\ell}}{\chi_*}\right) \frac{1}{P^\mathrm{tot}_{gg}(\mathbf{K}-\frac{\boldsymbol{\ell}}{\chi_*})C^\mathrm{tot}_\ell} \nonumber \\ &
\hspace{2em}\times\frac{1}{A^{\omega\omega}A^{uu} -A^{\omega u}A^{u\omega}} \left[ A^{uu} \bar{B}^{*\omega gT}-A^{u\omega}\bar{B}^{*ugT}\right],
\end{align}
where
\begin{align}
A^{XY}(\mathbf{K}) = \int\frac{\mathrm{d}^2\ell}{(2\pi)^2} \frac{\bar{B}^{XgT} \bar{B}^{* YgT}}{P(\mathbf{K}-\frac{\boldsymbol{\ell}}{\chi_*})C_\ell},
\end{align}
and $\bar{B}^{XgT}= B^{XgT}/P^{XX}$. Note that the vector power spectrum in the estimator completely cancels with equivalent terms implicit in the bispectra, so the estimator is independent of the shape of the velocity power spectrum.

\begin{figure*}
  \centering
  \includegraphics[width=.8\textwidth]{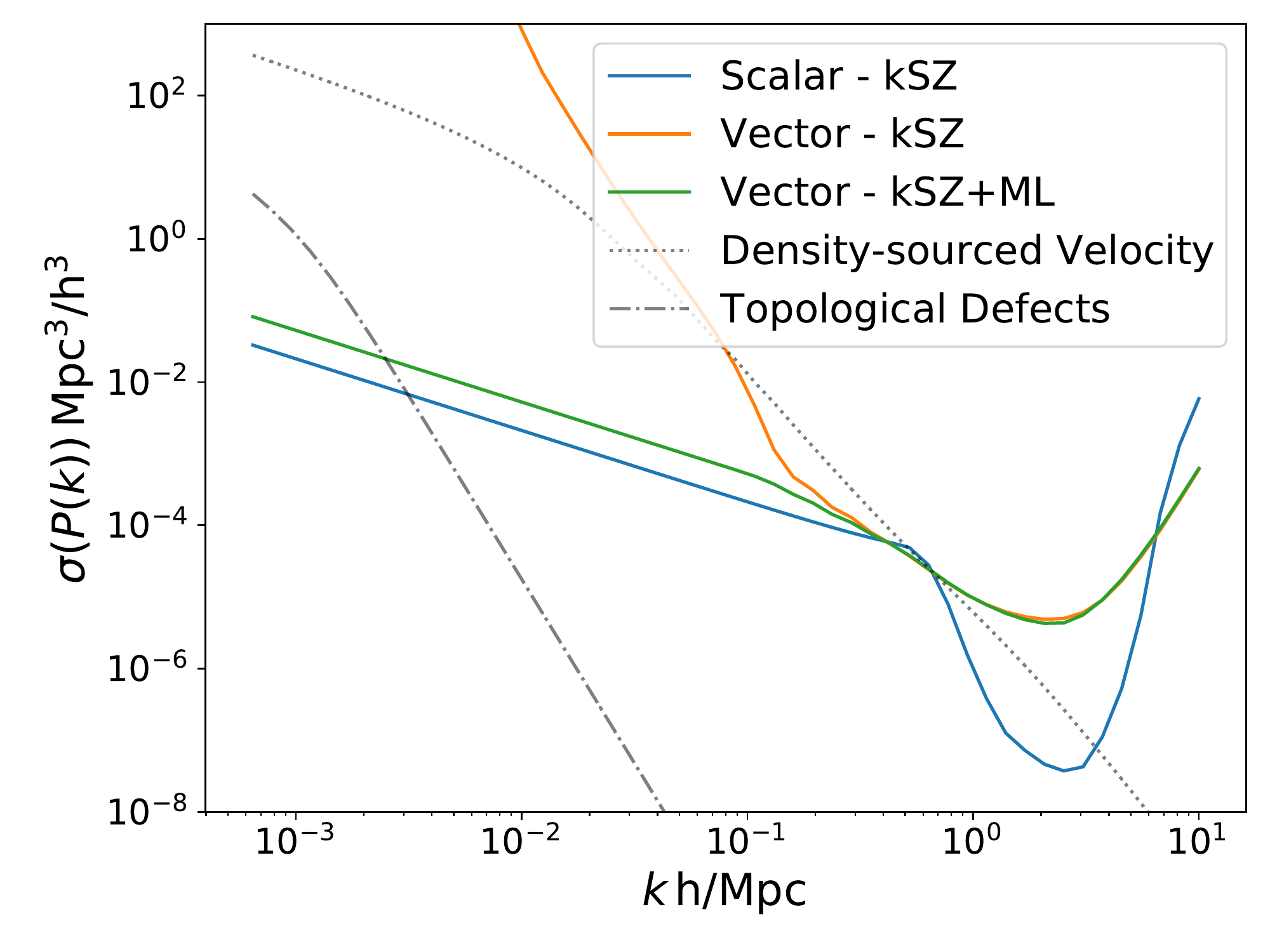} 
\caption{ A Fisher forecast for how well the DESI and CMB-S4 experiments can constrain the vorticity power spectrum. We show the constraints obtained using the kinetic Sunyaev-Zel'dovich effect alone and combining it with the moving lens effect. The joint analysis of these two effects allows the vorticity modes to be constrained independently of the scalar velocity modes. For comparison we show the constraints obtainable on the scalar velocity power spectrum,  the expected $\Lambda$CDM scalar velocity power spectrum (dotted) and the topological defect spectrum given in Eq. \ref{eq:top_def}. We use a logarithimic k-binning of width $\Delta \ln k=0.2$ }
\label{fig:reconstructedPk}
\end{figure*}

\section{Forecast constraints}\label{sec:constraints}
Using the formalism from \citet{smith2018ksz} we investigate the constraining power on this method. 
Specifically the noise power spectrum per mode for the standard estimator is given by
\begin{align}
N^\mathrm{kSZ-scalar}(\mathbf{K}) = \frac{1}{A^{uu}(\mathbf{K})},
\end{align}
and for the orthogonal vorticity estimator
\begin{align}
N^\mathrm{vorticity}(\mathbf{K}) = \frac{A^{uu}(\mathbf{K})}{A^{\omega\omega}(\mathbf{K})A^{uu}(\mathbf{K})-A^{u\omega}(\mathbf{K})A^{\omega u}(\mathbf{K})},
\end{align}
-- note that the noise is highly anisotropic. Then the uncertainity on the signal power spectrum, with an inverse noise weighting is
\begin{align}\label{eq:Pk_noise}
\frac{1}{\sigma^2(P^{XX}(K))} = V\int\frac{K^2\mathrm{d}K\sin\theta\mathrm{d}\theta_K\mathrm{d}\phi_K}{(2\pi)^{3}}\frac{1}{N^{XX}(\mathbf{K})}.
\end{align}
We consider how well a spectroscopic galaxy survey, like DESI \citep{DESI_2016}, in combination with a next generation CMB survey, such as CMB-S4 \citep{CMBS4_2016}, can measure these modes. For the DESI-like survey we use the experimental setup from \citet{smith2018ksz}
\citep[also see][]{Coulton_2020}, thus we assume a mean redshift of $z=0.75$,
a survey volume of  116~$(h^{-1}{\rm Gpc})^3$,
a galaxy number density of $1.7\times 10^{-4}$ Mpc$^{-3}$ and a galaxy bias of $b_g=1.51$. We use the CMB-S4 configuration described in \citet{CMBs4_2019} and compute the noise curves after component separation with the ILC method \citep{Tegmark_1996}. 

First we consider what can be learnt from only studying the kSZ anisotropies. The expected constraining power is shown in Fig.~\ref{fig:reconstructedPk}. Enforcing that the vector estimator has no response to the standard, scalar kSZ contribution comes at a huge noise cost. As stated earlier, this condition is necessary as the scalar modes are expected to be significantly larger than any vector contribution and without it vorticity measurements can be biased. The reason for the large noise cost can be seen from Eq.~\ref{eq:kSZ_allSources} -- whilst in principle the two sources have different angular dependencies and so could be separated, the region of phase space where there are purely scalar or vector sources is vanishing. Thus the vector modes constraints come from modes that are close to perpendicular to the line of sight.  If the scalar perturbations have an anisotropic power spectrum of the form $\sim\sin\theta_K/\cos\theta_K P_{uu}(|K|)$, their kSZ signature from this region of phase space would be highly similar to the vector modes. Our estimator makes no assumptions on the properties of the scalar perturbations and the large noise cost arises from accounting for the possibility of anisotropic scalar modes. 

For sufficiently low noise measurements it is possible to disentangle the two contributions with kSZ alone. Though a further complication arises as this separation requires modelling the bispectrum beyond the squeezed limit and, whilst the squeezed limit can be modeled with high accuracy \citep{Giri_2022}, accurately modeling the full bispectrum can be challenging. The ML-effect on its own can similarly be used to constrain vorticity modes, however it suffers the same issues and has even larger noise than kSZ based estimator (similar to that found for scalar velocity modes \citep{Hotinli_2019}). 

To alleviate these issues we consider using a combined estimator from the kSZ and ML effects. The ML and kSZ effects are sensitive to orthogonal velocities which is the precisely what is required to completely break the degeneracy.  The result of the combined estimator is shown in Fig.~\ref{fig:reconstructedPk}. The combined estimator provides powerful constraints on the large scale vorticity that are comparable to those of the scalar velocity mode. Note that this is achieved despite the low constraining power of the ML estimator; the contribution of the estimator is removal of the possibility of anisotropic scalar modes - as these would lead to a very large ML signal.

\section{Conclusions}\label{sec:conclusions}
Large-scale, cosmic vorticity modes in the late time Universe are an interesting cosmological observable. In the standard cosmological model these modes should be absent. Thus, searching for these modes is a powerful consistency test of our model and a discovery of these modes would represent the detection of new physics. 

In this work we show that the kSZ and ML effects, in combination with a galaxy survey, are an ideal method to search for these signals. Using tomography with  galaxy surveys, we are able to separate vorticity modes from velocity perturbations from scalar, density modes. Tomography enables the differentiation of fluctuations in the velocity field that are parallel to the velocity vector (density-source velocity modes) from those perpendicular to the velocity vector, vorticity modes. This joint method leads to almost white noise up to the largest scales accessible with the surveys, allowing for powerful constraints on the largest scales. Through combining galaxy and CMB measurements in a quadratic estimator to reconstruct the large scales, the method should have fewer large scale systematic effects. Especially as the non-trivial parity of the signal \citep[see e.g.][]{smith2018ksz} provides an extra means to suppress systematic biases. 

As a demonstration of the power of this approach, we consider constraining topological defects, a well known source of vorticity. Within this model the vorticity power spectrum is given as
\begin{align}\label{eq:top_def}
P_{\omega\omega}(k) = \frac{14 (G\mu)^2}{k^3}\left(\frac{k\tau/12}{1+(k\tau/12)^{3.13}} \right)
\end{align}
where $G \mu$ characterizes the amplitude of the topical defects \citep{Urrestilla_2008a}. Using our formalism we find that we can constraint $(G \mu)^2\le$ 1.6 $\times 10^{-10}$. Whilst topological defects leave strong and distinct signatures on the CMB, so are constrained significantly better by CMB measurements \citep{Dunkley_2011,Lizarraga_2016,Planck2013_XXV,Planck2016_XIII}, constraints obtainable with this kSZ formalism are more than two orders of magnitude tighter than those obtainable redshift space distortions \citep{Bonvin_2018}. This approach provides comparable constraints to those from CMB curl lensing \citep{Namikawa_2013} and by probing vector modes on a redshift slice, rather than the integrated over the lensing kernel, is highly complementary. This highlights the potential of this approach to probe new regimes in the late-time Universe.

Here we have only discussed one way to separate these scalar and vorticity modes, by combing the kSZ and ML estimators, when in practice many methods would be explored to ensure robust results. These could include using direct measurements of the density perturbations, such as through clustering measurements, to remove the scalar velocity modes (that are linearly related to density perturbations). We focused on the combined method as it is always necessary to account for both effects simultaneously to avoid biases from scalar modes from the other effect. This is necessary as separating the two effects in CMB maps is not possible with standard component separation methods, as both effects produce anisotropies with the same spectrum -- the same as the primary temperature anisotropies. The key other possible contaminants to this measurement are the thermal-kinetic SZ effect \citep{Coulton_2020}, which can be removed due to its distinct frequency spectrum, and gravitational lensing, which can be removed with established delensing techniques \citep{Seljak_2004,Smith_2012,Green_2017}.

A second use of the methods outlined here is as a null-test for the measurement of scalar velocity modes. A key goal of upcoming kSZ measurements is to search for primordial non-Gaussianity \citep[e.g.][]{Munchmeyer_2018} and one of the biggest challenges for such analyses is mitigating systematic biases \citep{Pullen_2013,Rezaie_2021}. If we assume there is no new physics, then the vector-mode velocity estimator can be used as a null-test for these analyses. This is analogous to how curl lensing \citep{Namikawa_2012} is ubiquitously used to test for systematic effects in weak lensing analyses.

\newpage
\appendix
\section{Light cone geometry}\label{app:lightConeGeom}
In this work we focused on discussing this analysis in the `snapshot' geometry picture, however equivalent expressions can be derived for the `light cone' picture. From this perspective we no longer work in a periodic box at a single redshift but instead in terms of spherical harmonic coefficients on the full sky and account for a more complex line of sight. 

In \citet{Terrana_2017,Deutsch_2018} an optimal, quadratic estimator is derived for the remote dipole field, $v^\mathrm{{eff}}$, from galaxy data in redshifts bins, denoted by $\alpha$. This estimator is
\begin{align}
  &  \hat{v^{\mathrm{eff}}_{\alpha,\ell m}} =  \nonumber \\ & N_{\alpha\ell}\sum\limits_{\ell_1 m_1 \ell_2 m_2}(-1)^m\Gamma^\mathrm{kSZ}_{\ell_1 \ell_2\ell \alpha}
\begin{pmatrix}
\ell_1 & \ell_2 &\ell \\ m_1 & m_2 & -m
\end{pmatrix}
\frac{a^T_{\ell_1 m_1} \delta_{g,\ell_2 m_2}^\alpha}{C_\ell^{TT}C_{\alpha \ell_2}^{\delta_g\delta_g}}
\end{align}
where $ \hat{v^{\mathrm{eff}}_{\alpha,\ell m}} $ is the estimated dipole spherical harmonic coefficients in a given redshift bin, $N_{\alpha \ell}$ is a normalization constant, $\Gamma^\mathrm{kSZ}_{\ell_1 \ell_2\ell \alpha}$ is a coupling matrix which combined with the Wigner $3j$ symbol is the equivalent of the bispectrum  in Eqs.~\ref{eq:estimatorScalars}-\ref{eq:wieghts}, $a^T_{\ell_1 m_1}$ and $\delta_{g,\ell_2 m_2}^\alpha$ are the spherical harmonic coefficients of the CMB map and galaxy density map and $C^{XX}_\ell$ are the power spectra of those maps.
This can be thought of as an equivalent expression to Eq.~\ref{eq:estimatorScalars}.

The expected contribution to this estimator from a scalar mode is given by
\begin{align}
    v^{\mathrm{eff}}_{\alpha,\ell m} =& \int \mathrm{d}\chi W^\alpha(\chi)\int\frac{\mathrm{d}^3k}{(2\pi)^3} \frac{4\pi i^\ell}{2\ell+1}\Phi(\mathbf{k})Y^*_{\ell m}(\hat{k}) \nonumber \\ & \times
    \left[\ell j_{\ell-1}(k\chi)-(\ell+1)j_{\ell+1}(k\chi) \right]
\end{align}
where $j_\ell(x)$ are the spherical Bessel functions and $W^\alpha(\chi)$ defines the redshift bin. Computing the equivalent expression for vector modes gives
\begin{align}
    v^{\mathrm{eff}}_{\alpha,\ell m} =& \int \mathrm{d}\chi W^\alpha(\chi)\int\frac{\mathrm{d}^3k}{(2\pi)^3} \frac{4\pi i^{\ell+1}}{2\ell+1}\omega^s(\mathbf{k})_{-s}Y^*_{\ell m}(\hat{k}) \nonumber \\ & \times \sqrt{\frac{\ell(\ell+1)}{2}}
    \left[ j_{\ell-1}(k\chi)+j_{\ell+1}(k\chi) \right].
\end{align}
The key difference between the two contributions is in the spherical Bessel projections -- one plus the other minus -- that is equivalent to the $\sin$ and $\cos$ terms in the vector/scalar kSZ estimator. Thus the  scalar and vector modes can be differentiated by combining observations at a range of redshift bins with different weightings.

\section*{Acknowledgments}

The authors are grateful for fruitful discussions with Colin Hill, David Spergel, Emmanuel Schaan, Anthony Challinor and Oliver Philcox. This work was supported in part by World Premier International Research Center Initiative (WPI Initiative), MEXT, Japan, and JSPS KAKENHI Grants No.~JP20J22055, JP20H05850, JP20H05855, JP19H00677, and by Basic Research Grant (Super AI) of Institute for AI and Beyond of the University of Tokyo. 
KA is supported by JSPS Overseas Research Fellowships. 

\bibliographystyle{mnras}
\bibliography{references}

\end{document}